\documentclass{elsart}
\usepackage{natbib}
\def\etal{{\it et al.\,}}
\def\flux{{$\times$10$^{-8}$ photons cm$^{-2}$ s$^{-1}$ }}
\def\gr{$\gamma$-ray }
\def\deg{{$^{\circ}$}}

\begin{document}
\runauthor{Sreekumar et al}
\begin{frontmatter}
\title{GeV emission from the nearby radio galaxy Centaurus A}
\author[gsfc,usra]{P. Sreekumar} 
\author[gsfc]{D.L. Bertsch}
\author[gsfc]{R.C. Hartman}
\author[stanford]{P.L. Nolan}
\author[gsfc]{D.J. Thompson}

\address[gsfc]{Code 661, NASA/Goddard Space Flight Center}
\address[usra]{Universities Space Research Association}
\address[stanford]{W.W. Hansen Laboratory, Stanford University}

\begin{abstract}
EGRET has detected 67 sources associated with active galactic nuclei.
With the exception of radio galaxy Cen A, all are
classified as belonging to the blazar class of active galactic nuclei.
The cumulative exposure from multiple EGRET observations has provided
the first clear detection of Centaurus A.  Unlike the \gr blazars seen
by EGRET which are believed to exhibit near-alignment
of the central jet along the line-of-sight, Cen A provides the first
evidence for $>$100 MeV emission from a source with a confirmed
large-inclination jet. Although the high-energy emission represents a
lower luminosity than most EGRET blazars, with the advent of new
more sensitive instruments such as GLAST and VERITAS, the
detection of off-axis high-energy emission from more distant radio
galaxies (space density of radio galaxies being $\sim$10$^3$ times the
blazar density) is an exciting possibility.
\end{abstract}
\end{frontmatter}

\section{Introduction}
Observations in the 30--10000 MeV energy range by the high-energy \gr
telescope EGRET on board the Compton Observatory (CGRO) has shown the
presence of a class of \gr bright blazars.
Blazars are in general characterized by flat radio spectra, rapid time
variability at most wavelengths and typically emit the bulk of their
bolometric luminosity at \gr energies. 
The recently released 3rd EGRET catalog
\cite{hart} contains 271 \gr sources of which 68 are
known to be extragalactic (66 blazars, 1 radio galaxy (Cen A) and 1 normal
galaxy (LMC)). Thus almost $\frac {2}{3}^{rd}$ of the known
\gr sources remain unidentified. 
 
We report here the
results from the analysis of all available EGRET data ($\sim$ 10 weeks
of on-axis exposure) on the nearby
radio galaxy Centaurus A. 
Centaurus A, at a  distance of $\sim$3.5 Mpc \cite{hui}, is the
 closest
 active galactic nucleus. 
 The proximity of Cen A has made it the subject of numerous studies at
 many
 wavelengths from radio to TeV energies. Radio studies have shown the
 presence of a double-lobed radio morphology \cite{sch1}.
 In the past, the presence of an
 obscuring dust lane has prevented detailed optical studies of the
 central nucleus
 and the inner regions of the jet. However recent NICMOS observations
 \cite {sch2} show extended emission and a bright unresolved
 central nucleus which may have associated with it a small ($\sim$40 pc
 diameter) inclined disk.
 A one-sided X-ray jet is visible \cite{sch3} and is collimated in the direction of the giant radio lobes.
 At TeV energies Grindlay \etal \cite{gr} used an early non-imaging
 Cherenkov system to report the discovery of
 emission from Cen A during a period of overall high activity 
 at lower frequencies.
 More recent observations by more sensitive instruments (CANGAROO)
 provide only upper limits \cite{rowell}.

Earlier attempts to detect Cen A from individual EGRET observations (typically 2 
weeks long) were hampered by weak detection significance which
resulted in large positional uncertainty. The possibility of association
of the \gr excess with another likely candidate, BL Lac object 1312-423
(1.95\deg away from Cen A), could not be ruled out . The strong
Galactic diffuse emission and the larger uncertainties associated with
the diffuse model \cite{hunter,sree1},
also contributed to the difficulties in confirming the detection early
on in
the mission \cite {fichtel,nol}.

\section{Results}
A likelihood analysis \cite{mattox} shows a 6.5$\sigma$ detection
of a point source type excess, positionally coincident with Centaurus A.
The average $>$100 MeV flux is (13.6$\pm$2.5)\flux.
The nearby BL Lac object 1312-423 is well outside the 95$\%$ confidence
contour. Unlike the variability seen at lower energies \cite{miyazaki}, the emission
above 100 MeV appears steady. The lack of variability could arise from
the near threshold detection associated with the individual
observations.  The 30-10000 MeV photon spectrum is well
characterized by a single power law of index 2.40$\pm$0.28.
This is steeper than the average power law spectrum from \gr blazars
(2.15$\pm$0.04) \cite{muk} and also steeper than the observed extragalactic
\gr background (2.10$\pm$0.03) \cite{sree1}. A comparison of
the EGRET measurements with OSSE \cite{kinzer} and COMPTEL
\cite{steinle} data at keV and MeV energies
yields a smooth, continuous spectrum that appears to evolve from a power
law above 200 keV (index=1.97) and steepen gradually above 1 MeV. 
The inclusion of significant systematic errors in comparing results
from different instruments is small, as evidenced by the good
cross-comparison
of the single power-law Crab pulsar spectrum across these energy bands
\cite{ulmer}.
A systematic search for \gr emission from other nearby radio
galaxies/Seyferts yielded no significant detection. 
 
\section{Conclusions}
 EGRET localization and spectral measurements provide unique
 confirmation that the source detected by OSSE and COMPTEL is
 Cen A.  Unlike the error regions defined by OSSE and COMPTEL, the
 nearby XBL 1312-4221 ($\sim$2\deg away from Cen A) lies well outside
 the EGRET 99$\%$ confidence contour. The consistency of the spectrum
 going from 50 keV to 1 GeV argues favorably for emission from
 a single source coincident with Cen A.
   
 A detailed analysis of EGRET archival data shows $>$100 MeV
 emission from only the nearest radio galaxy (Cen A). The low \gr
 luminosity of Cen A ($\sim$10$^{41}$ ergs/s, about 10$^5$ times smaller
 than the typical \gr blazar) if typical of this source class, provides
 the most likely explanation for the non-detection of more distant
 members of this source class.
	    
If Cen A is indeed a misaligned blazar \cite{bailey}
this provides
new evidence for $>$100 MeV emission from radio-loud AGN
with jets at large
inclination angles ($>$60\deg).  This is contrary to the model of Skibo, 
Dermer \& Kinzer \cite{sk} which suggests a significant cut off 
in the spectrum around a few MeV. Assuming a unification model 
for AGN, and increasing
high-energy emission with decreasing inclination angles
\cite{dermer}, the detection of more distant radio-loud
AGN with
intermediate jet inclination angles can be expected. This
provides a new
extragalactic source class for future high-energy
experiments such as
GLAST and VERITAS. Though the intrinsic luminosity is lower
than other
on-axis sources, the significantly larger space density of
radio-loud FR-I
sources, points to a new unresolved source class that could contribute
to the extragalactic \gr background around 1 MeV \cite{steinle}.
If the mean \gr spectrum of this new source class is harder than
the power-law spectral index of 2.40$\pm$0.28 observed for Cen A, then
significant contributions from this new source class are also expected
above 10 MeV.
A more detailed discussion including constraints on some theoretical 
models is provided elsewhere \cite{sree2}.

\end{document}